\documentclass[prb,twocolumn,preprintnumbers,showpacs,amsmath,amssymb]{revtex4}
\usepackage{color}
\usepackage{graphics}
\usepackage{dcolumn}
\usepackage{bm}

\renewcommand\textfraction 0
\renewcommand\topfraction 1
\renewcommand\bottomfraction 1
\begin{document}
\title{Tuning of magnetic and electronic states by control of oxygen content in lanthanum strontium
cobaltites} \author{S. Kolesnik, B. Dabrowski,  J. Mais, M. Majjiga, O Chmaissem} \affiliation{Department of Physics, Northern Illinois University, DeKalb, IL 60115}
\author{A. Baszczuk} \affiliation{INTiBS, PAN, Wroclaw, Poland}
\author{J. D. Jorgensen} \affiliation{Materials Science Division, Argonne National Laboratory, Argonne, IL 60439}

\date{\today}
\begin{abstract}
We report on the magnetic, resistive, and structural studies of perovskite La$_{1/3}$Sr$_{2/3}$CoO$_{3-\delta}$. By using the relation of synthesis
temperature and oxygen partial pressure to oxygen stoichiometry obtained from thermogravimetric analysis, we have synthesized a series of samples
with precisely controlled $\delta=0.00-0.49$. These samples show three structural phases at $\delta=0.00-0.15$, $\approx0.25$, $\approx0.5$,  and
two-phase behavior for other oxygen contents. The stoichiometric material with $\delta=0.00$ is a cubic ferromagnetic metal with the Curie
temperature $T_{\rm C}=274$ K. The increase of $\delta$ to 0.15 is followed by a linear decrease of $T_{\rm C}$ to $\approx$ 160 K and a
metal-insulator transition near the boundary of the cubic structure range. Further increase of $\delta$ results in formation of a tetragonal
$2a_p\times 2a_p \times 4a_p$ phase for $\delta\approx 0.25$ and a brownmillerite phase for $\delta\approx0.5$. At low temperatures, these are weak
ferromagnetic insulators (canted antiferromagnets) with magnetic transitions at $T_{\rm m}\approx230$ and 120 K, respectively.  At higher
temperatures, the $2a_p\times 2a_p \times 4a_p$ phase is $G$-type  antiferromagnetic between 230 K and $\approx$360 K. Low temperature magnetic
properties of this system for $\delta<1/3$ can be described in terms of a mixture of Co$^{3+}$ ions in the low-spin state and Co$^{4+}$ ions in the
intermediate-spin state and a possible spin transition of Co$^{3+}$ to the intermediate-spin state above $T_{\rm C}$. For $\delta>1/3$, there appears
to be a combination of Co$^{2+}$ and Co$^{3+}$ ions, both in the high-spin state with dominating antiferromagnetic interactions.

\end{abstract}


\maketitle

\section{Introduction}

Strontium substituted lanthanum cobaltites with a general formula La$_{1-x}$Sr$_x$CoO$_{3-\delta}$ have attracted much interest owing to
their possible applications in the field of solid oxide fuel cells \cite{Tedmon69,Goodenough65} as fast ion conducting materials and high
temperature oxygen separation membranes\cite{Kovalevsky98,Kharton99}. It has also been reported that substitution of Sr$^{+2}$ in LaCoO$_3$
results in a remarkable change in the crystal structure\cite {Yakel55,Mineshige96}, a metal-insulator transition,\cite{Mineshige96} and
ferromagnetic behavior. \cite{Senaris95} The structure of this compound with Sr substitution varying from 0 to 1 has been the subject of
numerous investigations at room temperature.\cite{Mineshige96,Sunstrom98,James04,VanDoorn00} A simple cubic structure has been reported by
most of the authors for samples with $x > 0.5$.\cite{Mineshige96,Sunstrom98,James04,VanDoorn00} Sunstrom et al.\cite{ Sunstrom98} prepared
the series of La$_{1-x}$Sr$_x$CoO$_{3-\delta}$ samples ($0.5 \leqslant x \leqslant 0.9$) using the Pechini gel technique and observed that
Sr-rich samples ($x > 0.7$) take a brownmillerite structure prior to oxidation and a cubic perovskite structure after being treated with
sodium hypobromite. Van Doorn and Burggraaf proposed an $a_p \times a_p \times 2a_p$ superstructure \cite{VanDoorn00} with distinct
microdomains in La$_{0.3}$Sr$_{0.7}$CoO$_{2.82}$. A few other regions without the superstructure (normal regions) have also been identified
and found to be fully oxygen-stoichiometric ($\delta= 0$), while the ones with the superstructure were found to have an oxygen content
$\delta= 0.5$.

Due to mixed valence of Co ions, in addition to oxygen nonstoichiometry and fast ion conducting properties, La$_{1-x}$Sr$_x$CoO$_{3}$
demonstrates a rich magneto-electronic phase diagram. One end-member of this family, LaCoO$_{3}$, is a nonmagnetic insulator with low-spin
Co$^{3+}$ ($t_{2g}^6$:$S=0$) at low temperatures, which undergoes a spin-state transition at around 100 K to the intermediate-spin
Co$^{3+}$ ($t_{2g}^5e_g^1$:$S=1$). \cite{Bhide72} The same transition has also been reported for small Sr concentrations $x \leqslant 0.15$
in La$_{1-x}$Sr$_x$CoO$_{3}$. \cite{Senaris95} Low Sr substitution ($x \leqslant 0.30$) leads to segregation of this material into
hole-rich ferromagnetic clusters in nonmagnetic matrix, similar to LaCoO$_{3}$. \cite{Senaris95} Material in this doping region behaves as
``cluster glass'' showing both spin-glass and ferromagnetic properties. \cite{Senaris95,Wu03} Increasing Sr concentration results in
coalescence of the clusters and the material becomes a ferromagnetic metal for $x \geqslant 0.3$. With an increase of $x$, pronounced
deviation (deficiency) from the ideal oxygen stoichiometry can be noticed for air synthesized samples. High-pressure oxygen ($\approx 2600$
bar) synthesis is required to obtain nearly stoichiometric material for the other end member of the family SrCoO$_{3-\delta}$
($\delta\approx 0.05$). \cite{Taguchi79} This material is a simple cubic perovskite metallic ferromagnet with Curie temperature $T_{\rm
C}=$~220 K (extrapolated to $\delta=0$). The electrochemical oxidation method was reported to give fully stoichiometric SrCoO$_{3}$.
\cite{Bezdicka93} This sample is also a metallic ferromagnet with $T_{\rm C}=$~280 K (determined from measurements in the magnetic field of
2 T).

Our work here describes a study of the effect of oxygen non-stoichiometry on the structure and magnetic properties of the
La$_{1/3}$Sr$_{2/3}$CoO$_{3-\delta}$ compound. A detailed description of the synthesis of La$_{1/3}$Sr$_{2/3}$CoO$_{3-\delta}$ with $0
\leqslant \delta \leqslant 0.49$ and precise determination of relation of oxygen content to the structure and physical properties have been
made. Our results clarify discrepant reports concerning properties of La$_{1-x}$Sr$_x$CoO$_{3-\delta}$ compounds for $x > 0.5$.

\section{SYNTHESIS AND EXPERIMENTAL TECHNIQUES}

\begin{table*}[htb]
\caption{Oxygen deficiencies, synthesis conditions, observed crystal structures, (c - cubic, 224 - $2a_p \times 2a_p \times 4a_p$
superstructure, bm - brownmillerite), reliability factors from the Rietveld refinements, and Co formal valence for
La$_{1/3}$Sr$_{2/3}$CoO$_{3-\delta}$ samples. The samples were synthesized either on a thermobalance (TGA), in a regular furnace or in a
high-pressure (h.p.) furnace.}
\begin{tabular}{cccccccccc}
\\
\hline \hline
\multicolumn{1}{c}{Oxygen} & \multicolumn{4}{c}{Synthesis conditions} & \multicolumn{1}{c}{Crystal} & \multicolumn{3}{c}{Reliability factors}& \multicolumn{1}{c}{Cobalt} \\
\cline{2-5} \cline{7-9}
deficiency $\delta$ & Facility & Atmosphere & T ($^{\circ}$C) & Cooling & structure & $R_{\rm p} (\%)$ & $R_{\rm wp} (\%)$ & $\chi^2$ & formal valence \\
\hline
0   &   h.p. furnace    &   250 bar O$_2$   &   500 &   slow    &    c  &   2.1 &   2.81    &   1.708   &    3.67\\
0.01    &   TGA &   100\% O$_2$ &   1100    &   slow    &    c  &   2.79    &   3.58    &   1.38    &    3.65\\
0.02    &   TGA &   21\% O$_2$/Ar   &   1100    &   slow    &    c  &   2.67    &   3.7 &   2.4 &    3.63\\
0.04    &   TGA &   1\% O$_2$/A &   1100    &   slow    &    c  &   2.24    &   3.22    &   2.45    &    3.59\\
0.07    &   furnace &   air &   1120    &   fast    &    c  &   2.37    &   3.31    &   2.147   &    3.53\\
0.08    &   furnace &   air &   400 &   quench in LN$_2$    &    c  &   2.34    &   3.02    &   1.67    &    3.51\\
0.15    &   furnace &   air &   600 &   quench in LN$_2$    &    c  &   2.07    &   2.77    &   1.609   &    3.37\\
0.22    &   furnace &   air &   800 &   quench in LN$_2$    &    c+224  &   1.93    &   2.48    &   1.56    &    3.23\\
0.25    &   furnace &   air &   860 &   quench in LN$_2$    &   224 &   1.77    &   2.25    &   1.43    &    3.17\\
0.28    &   furnace &   air &   930 &   quench in LN$_2$    &   224+bm  &   2.3 &   2.89    &   1.36    &   3.11\\
0.33    &   furnace &   air &   1080    &   quench in LN$_2$    &    224+bm     &   1.94    &   2.45    &   1.27    &   3.00\\
0.42    &   furnace &   Ar  &   900 &   fast    &    bm+224     &   2.63    &   3.39    &   1.96    &    2.83\\
0.47    &   furnace &   Ar  &   1000    &   fast    &    bm     &   1.95    &   2.5 &   1.42    &    2.73\\
0.49    &   TGA &   Ar  &   1100    &   fast    &    bm     &       &       &       &    2.69\\
\hline \hline\\
 \end{tabular}
 \end{table*}
The conventional solid state reaction was used to prepare single-phase samples of La$_{1/3}$Sr$_{2/3}$CoO$_{3-\delta}$. Appropriate molar
ratios of La$_2$O$_3$, SrCO$_3$ and Co$_3$O$_4$ were mixed and repetitively finely ground and fired in air at temperatures between
800-1000$^{\circ}$C. The sample was then pressed into pellets and fired at 1100$^{\circ}$C. Final sintering was done at 1120$^{\circ}$C to
obtain a dense, single phase sample. In each case sample was held for 12 hours at the specified temperature. In the next stage, multiple
batches of the sample each differing in the oxygen stoichiometry were obtained by additionally annealing the samples in high-pressure
oxygen, argon, or air followed by quenching in liquid nitrogen. Thermogravimetric analysis was used to determine the oxygen content in each
of these samples. The sample with highest oxygen content ($\delta= 0.00$) was obtained by annealing as-made sample under high-pressure
oxygen (250 atm) at 500$^{\circ}$C for several hours followed by very slow cooling to room temperature (0.1$^{\circ}$C/min). Annealing in
air under atmospheric pressure produced La$_{1/3}$Sr$_{2/3}$CoO$_{2.98}$ ($\delta= 0.02$).  The sample with $\delta= 0.49$ was obtained by
heating La$_{1/3}$Sr$_{2/3}$CoO$_{2.98}$ sample in pure Ar (99.9999\%) at 1080$^{\circ}$C and then quickly cooling it to room temperature.
The samples with intermediate oxygen deficiencies $\delta=$0.08-0.33 were obtained by annealing the as-made sample ($\delta= 0.07$) in air
at temperatures in the range 400-1080$^{\circ}$C and then rapidly quenching in liquid nitrogen. The synthesis conditions and respective
oxygen deficiencies are shown in Table I.

 The x-ray patterns of these samples were obtained by performing x-ray
diffraction at room temperature using a Rigaku Powder Diffractometer with Cu K$\alpha$ radiation. Data were collected for angles $2\theta$
ranging from 20 to 90 degrees with a step size of 0.02 and step time of 2.4 second. Time-of-flight neutron powder diffraction data was
collected at room temperature on the Special Environment Powder Diffractometer (SEPD)\cite{Jorgensen89} at the Intense Pulsed Neutron
Source (IPNS) at Argonne National Laboratory. High-resolution backscattering data (detector bank 1, 2$\theta$ = 144.85$^{\circ}$) were
analyzed using the Rietveld method with the GSAS (EXPGUI) suite.\cite{GSAS85} Thermogravimetric analysis with a Cahn TG 171 thermobalance
was used to determine the oxygen content. These experiments were conducted at temperatures 25-1100$^{\circ}$C in Ar, 1\% O$_2$/Ar, 21\%
O$_2$/Ar, and pure oxygen, and provided additional samples with well-defined oxygen deficiency $\delta=0.49$, 0.04, 0.02, and 0.01,
respectively. The oxygen content was confirmed with the accuracy $\delta \pm0.01$ by reduction of the $\delta=0.49$ sample in 1\%H$_2$/Ar
atmosphere on the thermobalance. The material decomposes in this atmosphere to binary oxides La$_{2}$O$_{3}$, SrO and elemental Co. A
Quantum Design MPMS system equipped with a 70 kOe superconducting magnet was used to measure dc magnetization and ac susceptibility. The
resistivity was measured with a Quantum Design PPMS system model 6000 using a standard four-point technique.

\section{Results and discussion}

\subsection{Structural properties}

\begin{figure}[!]
 \resizebox{8.5cm}{!}{\includegraphics{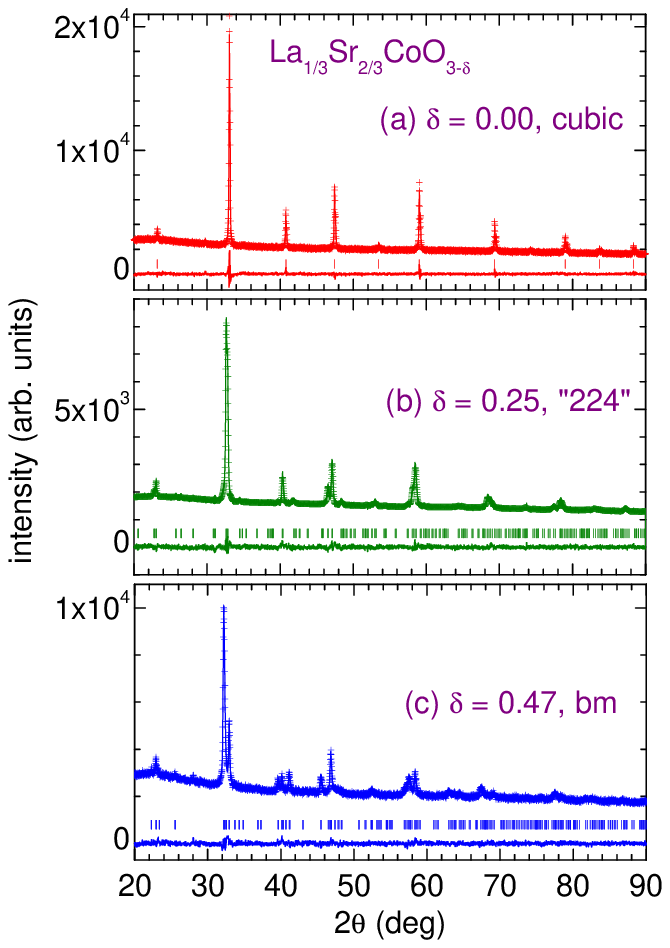}}
 \resizebox{8.5cm}{!}{\includegraphics{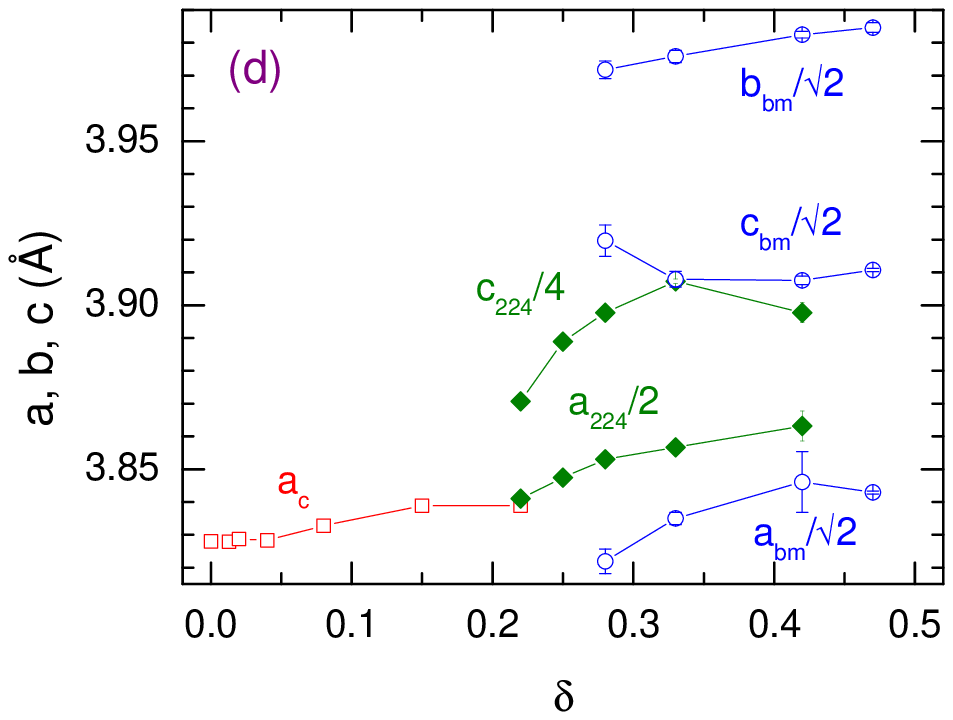}}
\caption{\label{LSCstruct} (color online) (a-c) X-ray diffraction patterns for La$_{1/3}$Sr$_{2/3}$CoO$_{3-\delta}$ samples with different
oxygen deficiencies $\delta$. Experimental data  points are presented as crosses. The continuous lines are the refined patterns and
differences between the data and the refined patterns. Intensity peak positions are marked as short vertical lines. (d) Lattice parameters
as a function of $\delta$ for three observed crystal structures: cubic (squares), ``224'' superstructure (diamonds), and brownmillerite (bm
- circles) are presented.} % The shaded areas represent the range of $\delta$ where two phases are observed.}
\end{figure}
Thermogravimetric measurements for various samples quenched in liquid nitrogen and fast cooled to room temperature indicate that the oxygen
non-stoichiometry, $\delta$, is strongly controlled by the atmosphere maintained during the final firing. In our experiments, we were able
to obtain La$_{1/3}$Sr$_{2/3}$CoO$_{3-\delta}$ samples with $\delta$ varying from 0 to $\sim$0.5. Refinements of the diffraction patterns
show that La$_{1/3}$Sr$_{2/3}$CoO$_{3-\delta}$ forms three single-phase compounds with cubic, ``224'' superstructure, and brownmillerite
crystal structures for $\delta$=0, 0.25, and 0.49, respectively. The x-ray diffraction patterns of selected samples are shown in
Fig.~\ref{LSCstruct}. We observe that samples with $0 \leqslant \delta \leqslant 0.15$ show single-phase x-ray diffraction patterns. These
could be indexed on simple cubic ($Pm$-$3m$) perovskite structure. They do not show any diffraction peaks that could be attributed to the
superstructure phase for this level of oxygen vacancies. The lattice parameter $a$ obtained from Rietveld refinements varies from 3.8280(1)
\AA~for $\delta=0$ to 3.8389(1) \AA~for $\delta=0.15$ which is the last member of the solid solution. The formal average charge of cobalt
for the samples in this region decreases from 3.667 to 3.367. Splitting and broadening of the diffraction peaks (Fig.~\ref{LSCstruct}) was
observed for samples with oxygen deficient compositions of $0.15<\delta<0.25$. These samples were found to contain two crystallographic
phases with $\delta=$0.15 and $\delta=$0.25. The x-ray diffraction profile of La$_{1/3}$Sr$_{2/3}$CoO$_{2.75}$ revealed a nearly
single-phase sample that could be indexed using a tetragonal $2a_p \times 2a_p \times 4a_p$ (``224'') superstructure, where $a_p$ is the
cell parameter of the cubic perovskite. However, since we were not able to observe several of the diffraction peaks unique for this
superstructure, we have prepared additional large-size sample with oxygen content $\delta=$0.28 for detailed investigation using neutron
diffraction, which is described in the following subsection.

The x-ray diffraction pattern for La$_{1/3}$Sr$_{2/3}$CoO$_{2.53}$ indicate a single-phase sample. Rietveld refinement proves that sample
with oxygen deficiency $\delta = 0.47$ crystallizes with the brownmillerite structure  (space group $Icmm$) with alternate layers of
[CoO$_6$] octahedra and oxygen deficient layers of [CoO$_4$] tetrahedra with randomly mixed La$^{3+}$/Sr$^{2+}$ cations occupying the same
crystallographic site. The oxygen vacancies in these [CoO$_4$] layers were found to be ordered in such a way that they formed tetrahedra
with corners linked in chains running along the [001] direction. \cite{VanDoorn00} For the samples with $0.25<\delta<0.47$, two
crystallographic phases are observed, namely the ``224'' superstructure and brownmillerite.

\subsection{Neutron powder diffraction of La$_{1/3}$Sr$_{2/3}$CoO$_{2.72}$}

\begin{figure}[!]
 \resizebox{8.5cm}{!}{\includegraphics{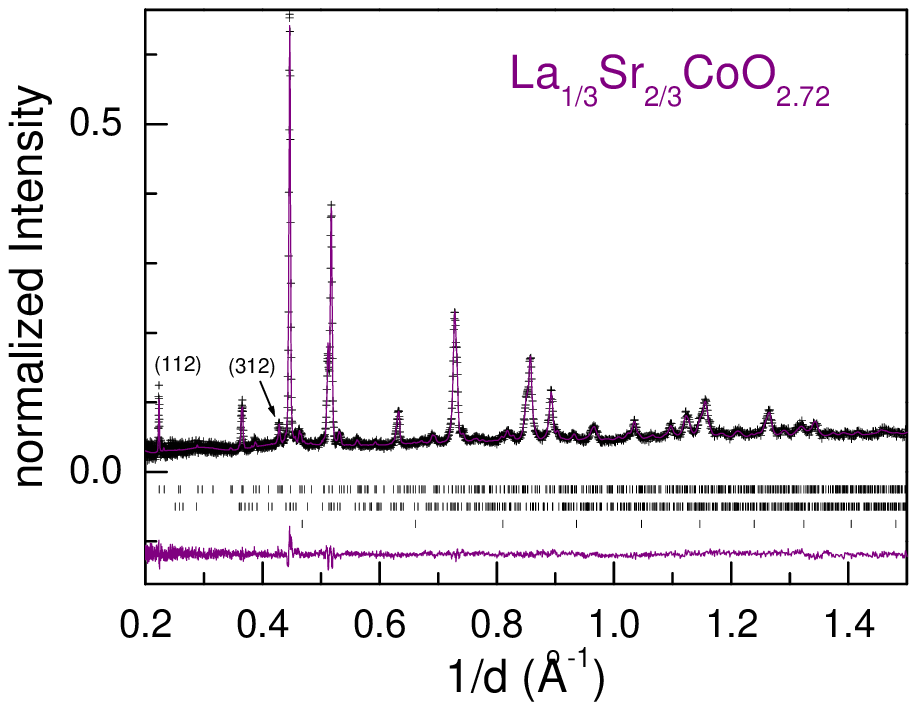}}
\caption{\label{LSCneut} (color online) Neutron diffraction pattern for La$_{1/3}$Sr$_{2/3}$CoO$_{2.72}$. Experimental data  points are
presented as crosses. The continuous lines are the refined patterns and differences between the data and the refined patterns. Intensity
peak positions are marked as short vertical lines for the ``224'' superstructure, brownmillerite, and vanadium can (from top to bottom
respectively).}
\end{figure}
The neutron powder diffraction (NPD) pattern of La$_{1/3}$Sr$_{2/3}$CoO$_{2.72}$ was successfully indexed with $I$-centered tetragonal unit cell
related to cubic perovskite subcell by $a \approx 2 a_p$ and $c \approx 4 a_p$. This kind of superstructure (called in following 224) was previously
reported for several $A$-site substituted cobalt perovskites $Ln_{1-x}$Sr$_{x}$CoO$_{3-\delta}$ with
$x=$0.67~\cite{Withers03,James04,Goossens04,Goossens05} and $x=$0.6~\cite{Istomin03,Istomin04}, where $Ln=$ Sm-Yb and Y. The origin of the
superstructure was rationalized by partial $Ln$/Sr ordering on three crystallographic $A$-sites (one of them exclusively occupied by $Ln$ and two
mixed) and a presence of oxygen vacancies located solely on the O2 sites. The initial model used for the crystal structure refinement of
La$_{1/3}$Sr$_{2/3}$CoO$_{2.72}$ was taken from Ref. \cite{Withers03}, with three disordered $A$-cation sites, two Co sites, and four oxygen sites
initially set as fully occupied. The refinement revealed that oxygen vacancies were located on the O2 sites, while the O1, O3 and O4 sites were fully
occupied. The refined amount of oxygen vacancies was smaller than the one determined by thermogravimetric analysis, probably because of a small
fraction of brownmillerite phase (refined to $\sim$2.5\% wt.) present in the sample. Similar discrepancies between refined amount of oxygen and that
obtained from TGA were observed previously for $Ln_{0.33}$Sr$_{0.67}$CoO$_{3-\delta}$ compounds with $Ln=$ Ho and Y.~\cite{ Goossens04,Goossens05} In
the following refinements the amount of oxygen vacancies was fixed at the TGA value. During the refinement it was found that calculated intensity of
some of the diffraction peaks (in particular the 112 and 312) in the room temperature profile of La$_{1/3}$Sr$_{2/3}$CoO$_{2.72}$ were always lower
than the measured intensities. The additional peak intensity could not be modelled by changes of the atomic structure, but rather, were due to
magnetic Bragg scattering from ordered magnetic moments. Similar enhancement of the peak intensities was observed below antiferromagnetic transition
temperature for Sr$_{0.67}$Y$_{0.33}$CoO$_{2.79}$~\cite{Goossens04} and Sr$_{0.7}$Dy$_{0.3}$CoO$_{2.62}$~\cite{Istomin04}. We have tested several
patterns of magnetic spin ordering for cobalt atoms and found that the best fit was observed using magnetic space group $I4/mmm'$. The magnetic
refinement revealed antiferromagnetic alignment along all three crystallographic directions corresponding to $G$-type magnetic structure found for
perovskite-type La$_{1-x}$Ca$_{x}$MnO$_{3}$ manganites~\cite{Wollan55}. The same $G$-type magnetic structure was observed recently for
Ho$_{0.33}$Sr$_{0.67}$CoO$_{2.76}$~\cite{Goossens05}, Ho$_{0.1}$Sr$_{0.9}$CoO$_{2.79}$~\cite{Goossens05},
Sr$_{0.67}$Y$_{0.33}$CoO$_{2.79}$~\cite{Goossens04},  and Sr$_{0.7}$Dy$_{0.3}$CoO$_{2.62}$~\cite{Istomin04} cobaltites with the 224 superstructure
~\cite{Goossens04,Goossens05,Istomin04}. The magnetic moments of cobalt atoms are directed along the $c$-axis. During refinement magnetic moments of
Co1 and Co2 were constrained to be equal and their refined values found at 1.62(7) $\mu_B$.  The neutron diffraction pattern of
La$_{1/3}$Sr$_{2/3}$CoO$_{2.72}$ and a difference between the measured and calculated patterns are shown in Fig.~\ref{LSCneut}. Crystallographic
parameters, refinement data, atomic positions, and thermal parameters are given in Table II.
\begin{table*}[htb]
\caption{Crystallographic parameters, atomic positions, thermal parameters, and reliability factors from the Rietveld refinements for
La$_{1/3}$Sr$_{2/3}$CoO$_{2.72}$.}
\begin{tabular}{ccccccc}
\\
\hline \hline
Atom & Site & $x$ & $y$& $z$ & B$_{iso}$ (\AA) & Occ. \\
\hline
Sr1/La1 &    $8g$   &   0   &   0.5 &   0.1384(6)   &   1.12(17)    &   0.667/ 0.333    \\
Sr2/La2     &   $4e$    &   0   &   0   &   0.6335(10)  &   1.13(37)    &   0.667/ 0.333    \\
Sr3/La3     &   $4e$    &   0   &   0   &   0.127l(8)   &   0.41(28)    &   0.667/ 0.333    \\
Co1     &   $8h$    &   0.2442(31)  &   0.2442(31)  &   0   &   1.30(31)    &   1   \\
Co2     &   $8f$    &   0.25    &   0.25    &   0.25    &   1.86(31)    &   1   \\
O1  &   $16m$   &    0.2514(12)     &   0.2514(12)  &   0.1207(4)   &   B$_{11}$= 5.13(34) B$_{22}$= 5.13(34)   &   1   \\
    &       &       &       &       &   B$_{33}$= 2.9(5) B$_{12}$ = -3.3(6)     &       \\
    &       &       &       &       &   B$_{13}$= 1.0(4) B$_{23}$ = 1.0(4)  &       \\
O2  &   $8i$    &    0.2632(63)     &   0   &   0   &   B$_{11}$=-8.16(337) B$_{22}$= -0.05(85)  &   0.441 \\
    &       &       &       &       &   B$_{33}$=-1.2(9) B$_{12}$ = 0   &       \\
    &       &       &       &       &   B$_{13}$=0 B$_{23}$ = 0     &       \\
O3  &   $8j$    &    0.2787(48)     &   0.5 &   0   &    B$_{11}$=15.16(216) B$_{22}$=4.98(103)     &   1   \\
    &       &       &       &       &   B$_{33}$=1.5(9) B$_{12}$ = 0   &       \\
    &       &       &       &       &   B$_{13}$=0 B$_{23}$=0   &       \\
O4  &   $16n$   &   0   &   0.2475(18)  &   0.2475(18)  &    B$_{11}$= 0.22(19) B$_{22}$=1.84(32)   &   1   \\
    &       &       &       &       &   B$_{33}$=5.3(5) B$_{12}$=0  &       \\
    &       &       &       &       &   B$_{13}$=0 B$_{23}$=-3.9(4)     &       \\
\hline
Space group: $I4m/m~m$   &       &       &       &       &       &       \\
$a$ = 7.7212(21) \AA  &       &       &       &       &       &       \\
$c$ = 15.613(4) \AA  &       &       &       &       &       &       \\
$V$ = 930.8(7) \AA$^3$  &       &       &       &       &       &       \\
\hline
$\chi^2$    &   1.612   &       &       &       &       &       \\
$R_{wp}$ (\%)   &   5.68    &       &       &       &       &       \\
$R_p$ (\%)  &   4.24    &       &       &       &       &       \\
\hline \hline\\
 \end{tabular}
 \end{table*}
An examination of the bond lengths shows (Table III) that the average Co1-O bond length is 1.9237~\AA~while Co2-O is considerably longer
(1.9600~\AA).

\begin{table}[htb]
\caption{Selected interatomic distances for La$_{1/3}$Sr$_{2/3}$CoO$_{2.72}$. }
\begin{tabular}{ccc}
\\
\hline \hline
\multicolumn{1}{c}{Bonds} & \multicolumn{1}{c}{Multiplication} & \multicolumn{1}{c}{Distance (\AA)} \\
\hline
Co1-O1    &   (*2)    &   1.887(7)    \\
Co1-O2    &   (*2)    &   1.893(23)   \\
Co1-O3    &   (*2)    &   1.991(27)   \\
$<$Co1-O$>$   &       &   1.9237  \\
Co2-O1    &   (*2)    &   2.017(6)    \\
Co2-O4    &   (*4)    &   1.9315(7)   \\
$<$Co2-O$>$    &       &   1.9600  \\
Sr1-O1    &   (*4)    &   2.7437(12)  \\
Sr1-O2    &   (*2)    &   2.831(32)   \\
Sr1-O3    &   (*2)    &   3.050(26)   \\
Sr1-O4    &   (*2)    &   2.570(18)   \\
Sr1-O4    &   (*2)    &   2.659(12)   \\
$<$Sr1-O$>$  &       &   2.7662  \\
Sr2-O1    &   (*4)    &   2.725(13)   \\
Sr2-O3    &   (*4)    &   2.696(26)   \\
Sr2-O4    &   (*4)    &   2.683(22)   \\
$<$Sr2-O$>$  &       &   2.7013  \\
Sr3-O1    &   (*4)    &   2.743(13)   \\
Sr3-O2    &   (*4)    &   2.84(4)     \\
Sr3-O4    &   (*4)    &   2.661(14)   \\
$<$Sr3-O$>$  &       &   2.748   \\
\hline \hline\\
 \end{tabular}
 \end{table}

Our results are similar to those found for others cobalt perovskites with 224 superstructure, for which charge ordering was suggested for Co1
(Co$^{3+}$) and Co2 (Co$^{4+}$)~\cite{James04,Goossens04}. Using disordered $A$-cation positions the refined average $<$Sr/La-O$>$ bond lengths were
found very similar for all three A-sites ranging from 2.7013 to 2.7662~\AA. These values are in very good agreement with those observed for other
randomly substituted strontium/lanthanum cobaltites with the mixed A-site position~\cite{Mineshige96,Sunstrom98}. To confirm this structure model,
that infers no $A$-site ordering, a part of the sample with oxygen deficiency $\delta=$0.28 was annealed under high-pressure oxygen at low
temperatures (180 bar at 430$^{\circ}$C, followed by slow cooling) in order to obtain fully oxygenated compound with $\delta=$0. The annealed sample
showed single-phase x-ray diffraction pattern which could be indexed using simple cubic ($Pm-3m$) perovskite structure similar to that previously
discussed for the $\delta=$0 sample obtained from 250 bar and 500$^{\circ}$C. The absence of superlattice peaks confirmed mixed occupancy of the
$A$-site by La and Sr for both samples with oxygen contents before ($\delta=$0.28) and after annealing ($\delta=$0). Our results suggest, that in
contrast to other cobaltites with smaller rare earth elements $Ln_{1-x}$Sr$_{x}$CoO$_{3-\delta}$ ($x=$0.67 and 0.7) exhibiting the $a \approx 2a_p$
and $c \approx 4a_p$ superstructure, there is no Sr/La ordering in La$_{1/3}$Sr$_{2/3}$CoO$_{2.72}$. This observation allows us to exclude the role
of $A$-site cation ordering in preferential ordering of oxygen vacancies on the O2 sites. Consequentially, oxygen vacancy ordering alone seems to be
the main reason for occurrence of the complex ``224'' superstructure. Similar conclusions were recently made by S. Malo and A. Maignan~\cite{Malo04}
who observed the ``224'' superstructure for the SrTi$_{0.1}$Co$_{0.9}$O$_{3-\delta}$ compound with exclusively Sr$^{2+}$ cations on the $A$-site. For
highly doped cobaltites $Ln_{1-x}A_{x}$CoO$_{3-\delta}$ ($x\geqslant$0.5, $Ln$ = rare earth, $A$ = alkaline earth) several perovskite-related
superstructures have been reported. For example, the $a_p \times a_p \times 2a_p$ superstructure was proposed for
La$_{0.3}$Sr$_{0.7}$CoO$_{2.82}$~\cite{VanDoorn00}, Ho$_{0.1}$Sr$_{0.9}$CoO$_{2.73}$~\cite{Streule04} and for $Ln_{1-x}$Ba$_{x}$CoO$_{3-\delta}$
compounds.~\cite{Maignan99} Clearly, this superstructure model can not account for reflections observed in the NPD pattern of our sample and could
not be used for La$_{1/3}$Sr$_{2/3}$CoO$_{2.72}$. The $Ln_{1-x}$Ba$_{x}$CoO$_{3-\delta}$ system revealed the $a_p \times 2a_p \times 2a_p$ and $3a_p
\times 3a_p \times 2a_p$ superstructures arising from vacancy ordering in [$Ln$O$_{\delta}$] layer~\cite{Maignan99,Kim00}.  Using these models with
larger supercells we were not able to index all observed diffraction peaks in the NPD pattern for La$_{1/3}$Sr$_{2/3}$CoO$_{2.72}$. For example,
reflections at $d$ = 1.78, 1.99, 2.33, 2.44, 4.47 are not accounted in the $a_p \times 2a_p \times 2a_p$ model while reflections at $d$ = 2.33, 2.44,
4.47 can not be indexed for the $3a_p \times 3a_p \times 2a_p$ superstructure. These peaks can be indexed only in the ``224'' superstructure [(312),
(310), and (112), respectively].

\subsection{Transport properties}

\begin{figure}[!]
 \resizebox{8.5cm}{!}{\includegraphics{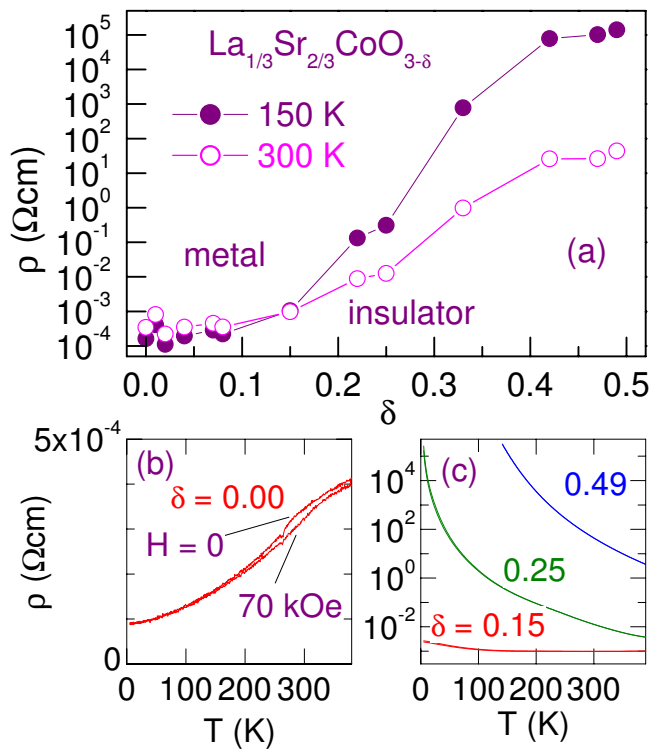}}  \caption{\label{LSCrho} (color online)
(a) Resistivity at $T=150$ and 300~K for La$_{1/3}$Sr$_{2/3}$CoO$_{3-\delta}$. (b),(c) Temperature dependencies of resistivity for selected
values of $\delta$.}
\end{figure}

Resistivity for several La$_{1/3}$Sr$_{2/3}$CoO$_{3-\delta}$ samples is presented in
Fig.~\ref{LSCrho}.
The temperature dependence of resistivity shows  metallic character for small oxygen deficiency ($0 \leqslant \delta \leqslant 0.08$). These samples
are metallic below and above the Curie temperature $T_{\rm C}$ and show a weak negative magnetoresistance around $T_{\rm C}$ [see:
Fig.~\ref{LSCrho}(b)]. An insulating character is observed for both single-phase and mixed-phase samples with $0.22 \leqslant \delta \leqslant 0.49$
[Fig.~\ref{LSCrho}(c)]. The crossover between these two types of behavior is observed for $\delta\approx0.15$ at the boundary of the cubic perovskite
structure [Fig.~\ref{LSCrho}(a)]. It appears that the metal/insulator transition in La$_{1/3}$Sr$_{2/3}$CoO$_{3-\delta}$ is controlled by both the
band filling (Co formal valence) and localization due to disorder (the oxygen vacancies $\delta$). For stoichiometric La$_{1-x}$Sr$_x$CoO$_{3}$
samples, the metal/insulator transition was observed at strontium substitution of $x\approx0.25$ in the rhombohedral structure. \cite{Mineshige96}
Since $x$ in fully oxygenated La$_{1-x}$Sr$_x$CoO$_{3}$ is equal to the fraction of Co$^{4+}$ ions, we can see that the metal/insulator transition
(by changing the oxygen content) in La$_{1/3}$Sr$_{2/3}$CoO$_{3-\delta}$ takes place for significantly larger fraction of Co$^{4+}$ ($\approx37$\%,
see: Table I) than for La$_{1-x}$Sr$_x$CoO$_{3}$ ($\approx25$\%) That indicates that the insulating state is enhanced by the disorder already in the
cubic structure for $\delta\approx0.15$ and fully quenched in ``224'' and brownmillerite phases with Co formal valences $\approx3.17$ and
$\approx2.7$, respectively.

\subsection{Magnetic properties}

\begin{figure}[!]
 \resizebox{8.5cm}{!}{\includegraphics{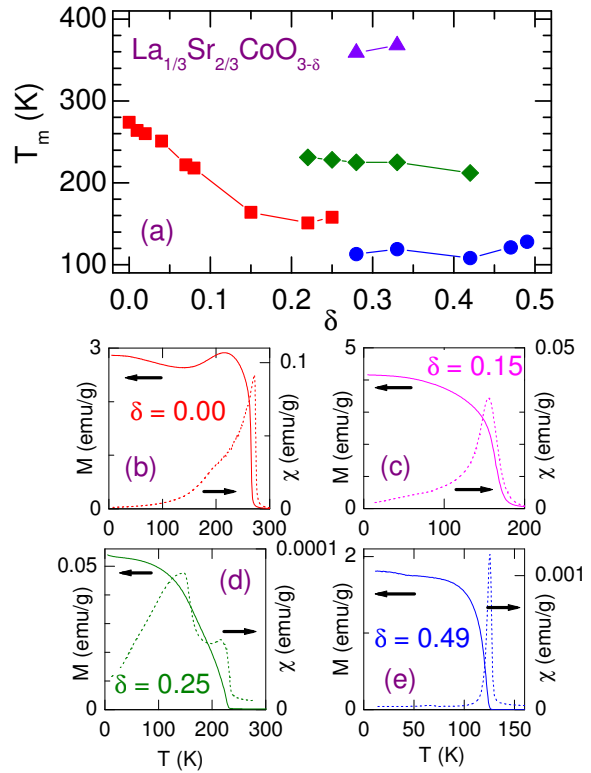}}
 \caption{\label{LSCTc} (color online)
(a) Magnetic transition temperatures for La$_{1/3}$Sr$_{2/3}$CoO$_{3-\delta}$. (b)-(e) Temperature dependencies of ac susceptibility (dashed lines,
right axis) and dc magnetization (``field cooled'' in 20 Oe, solid lines, left axis) for selected $\delta$. }
\end{figure}

The ac susceptibility and dc magnetization for several La$_{1/3}$Sr$_{2/3}$CoO$_{3-\delta}$ samples are presented in Fig.~\ref{LSCTc}(b)-(e).
The magnetic transition temperatures ($T_{\rm m}$) were determined from the dc magnetization curves $M(T)$ (measured on cooling in the magnetic field
of 20 Oe) as the temperatures of the maximum slope $-dM/dT$ [Fig.~\ref{LSCTc}(a)]. These temperatures also coincide with the peaks of the ac
susceptibility. The samples with cubic structure show ferromagnetic transitions with temperatures  linearly decreasing from 274 to 160 K for $\delta$
increasing from 0 to 0.15. The ``224'' ($\delta\approx 0.25$) and brownmillerite ($\delta\approx0.5$) phases show weak ferromagnetic transitions at
230 and 120 K, respectively. The samples that exhibit a mixture of two crystallographic phases also show two ferromagnetic transitions. Therefore, we
can find a one-to-one correlation between the crystal structures and ferromagnetic phases in La$_{1/3}$Sr$_{2/3}$CoO$_{3-\delta}$
[Fig.~\ref{LSCstruct}(b) and Fig.~\ref{LSCTc}(a)]. It is important to note that the two-phase nature of the samples is sometimes difficult to observe
in magnetization measurements alone. On the other hand, the ac susceptibility shows clear two transitions for such samples.

The magnetization in the intermediate temperature range shows broad antiferromagnetic transitions for most of the samples, for which the $2a_p\times
2a_p \times 4a_p$ is the majority phase ($\delta=0.28$ and 0.33). The transition temperatures for the $\delta=0.28$ sample (studied with NPD) and the
$\delta=0.33$ sample are 359 and 368 K, respectively. The small change of magnetization at the transition temperature is typical of the $G$-type
antiferromagnetic materials and usually is better visible in a higher magnetic field (10 kOe in this case).

\begin{figure}[!]
 \resizebox{8.5cm}{!}{\includegraphics{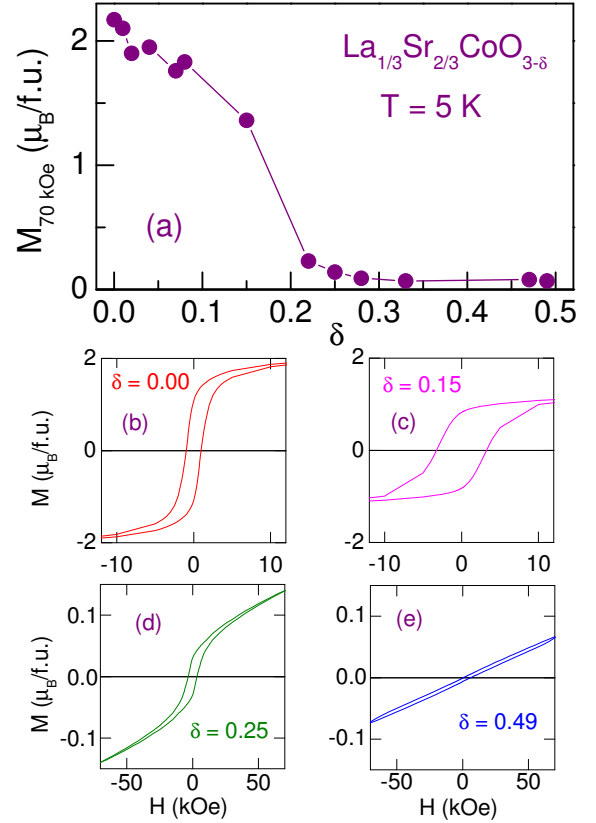}}
 \caption{\label{LSCmsat} (color online)
(a) High-field magnetization in $H=70$ kOe and $T=5$ K for La$_{1/3}$Sr$_{2/3}$CoO$_{3-\delta}$. (b)-(e) Magnetization curves for several values of
$\delta$ at 5 K.}
\end{figure}

The magnetization hysteresis curves for selected La$_{1/3}$Sr$_{2/3}$CoO$_{3-\delta}$ samples at $T=5$ K are presented in Fig.~\ref{LSCmsat}(b)-(e).
The high-field magnetization, determined from the magnetization curves, shows a clear saturation for $\delta \leq 0.15$. These samples demonstrate a
typical ferromagnetic behavior. Therefore, we refer to the high-field magnetization as to the saturation magnetization. The magnetic transition
temperatures are also equivalent to the ferromagnetic Curie temperatures ($T_{\rm m}\equiv T_{\rm C}$). The saturation magnetization linearly
decreases in this range of $\delta$ [Fig.~\ref{LSCmsat}(a)]. For higher $\delta$, the high-field magnetization drastically decreases and a linear
contribution to the M(H) curve is observed. These observations indicate a weak nature of ferromagnetism, probably associated with a canted
antiferromagnetic state for ``224'' and brownmillerite phases. Future neutron diffraction experiments are expected to reveal the actual type of
magnetic ordering at low temperatures in our samples.

The molar dc susceptibility $\chi_m=M/H$ in the temperature range 300-800 K was fitted to the
general Curie-Weiss formula:
\begin{equation}
 \chi_m = \chi_0 + (\mu_B N_A/3k_B)\mu_{\rm eff}^2/(T-\Theta),
\end{equation}
where $\chi_0$ is a temperature-independent background susceptibility, $N_A$ is the Avogadro constant, $k_B$ is the Boltzmann constant, $\Theta$ is
the paramagnetic Curie-Weiss temperature, $\mu_{\rm eff}$ is the effective paramagnetic moment. For the samples with smaller $\delta<0.15$ we used a
narrower fitting range 300-500~K because of the oxygen loss from the samples above 500~K during the measurements in the magnetometer under a low
pressure of helium gas. For the sample with $\delta=0.33$, which showed the antiferromagnetic transition at 368~K, the fitting range 430-800~K was
applied.

\begin{figure}[!]
 \resizebox{8.5cm}{!}{\includegraphics{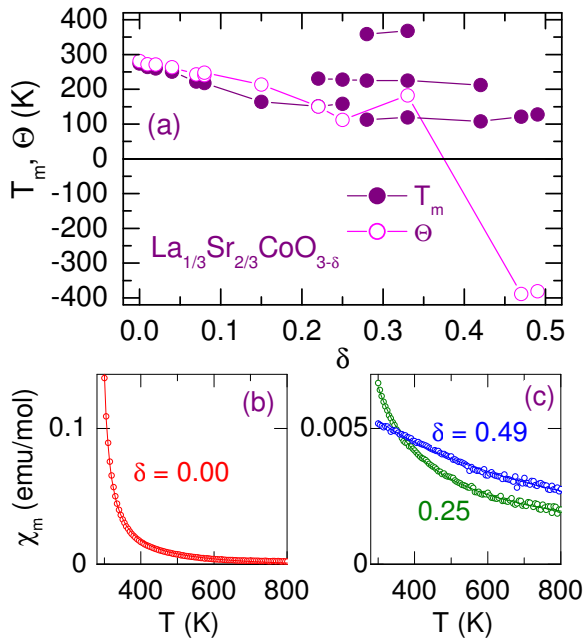}}
 \caption{\label{LSCtheta} (color online)
(a) The magnetic transition temperatures $T_{\rm m}$ (filled circles) and Curie-Weiss temperatures $\Theta$ (open circles) for
La$_{1/3}$Sr$_{2/3}$CoO$_{3-\delta}$. (b),(c) Molar susceptibility data for several values of $\delta$, measured in $H=1$ kOe. Lines are
fits of the Curie-Weiss formula to the susceptibility data.}
\end{figure}

The characteristic  temperatures $T_{\rm m}$ and $\Theta$ are presented in Fig.~\ref{LSCtheta}(a) for La$_{1/3}$Sr$_{2/3}$CoO$_{3-\delta}$. For small
values of $\delta\leqslant0.15$, $\Theta$ is positive and closely follows $T_{\rm C}$. For each two-phase sample with $\delta=0.22$-$0.33$, $\Theta$
is also positive, which is a resultant value for the two coexisting phases. For $\delta=0.47$ and 0.49, $\Theta$ becomes negative, which indicates a
dominating contribution of antiferromagnetic interactions between random cobalt spins in the paramagnetic state.

The analysis of the effective paramagnetic moment $\mu_{\rm eff}$ along with the saturation magnetization $M_{\rm sat}$ provides us with
information on the spin states of magnetic cobalt ions in La$_{1/3}$Sr$_{2/3}$CoO$_{3-\delta}$.  Formal valence of cobalt in this material
can be expressed as ($3+2/3-2\delta$)+, as listed in Table I.

It is worth noting that after increasing oxygen deficiency to $\delta>1/3$, the formal Co valence is less than 3+, which is unique to Sr
substituted rare-earth cobaltites. In this case, a mixture of Co$^{2+}$ and Co$^{3+}$ ions results in dominating antiferromagnetic
interactions and small spin canting with net weak ferromagnetic behavior. Large values of the effective paramagnetic moment in this oxygen
content range suggest high-spin configuration of both Co$^{2+}$ and Co$^{3+}$ ions (Co$^{2+}$HS+Co$^{3+}$HS). This configuration has been
found to satisfactorily describe the properties of Co$^{2+}$/Co$^{3+}$ containing $Ln$BaCo$_2$O$_{5}$. \cite{Suard00}

\begin{figure}[!]
 \resizebox{8.5cm}{!}{\includegraphics{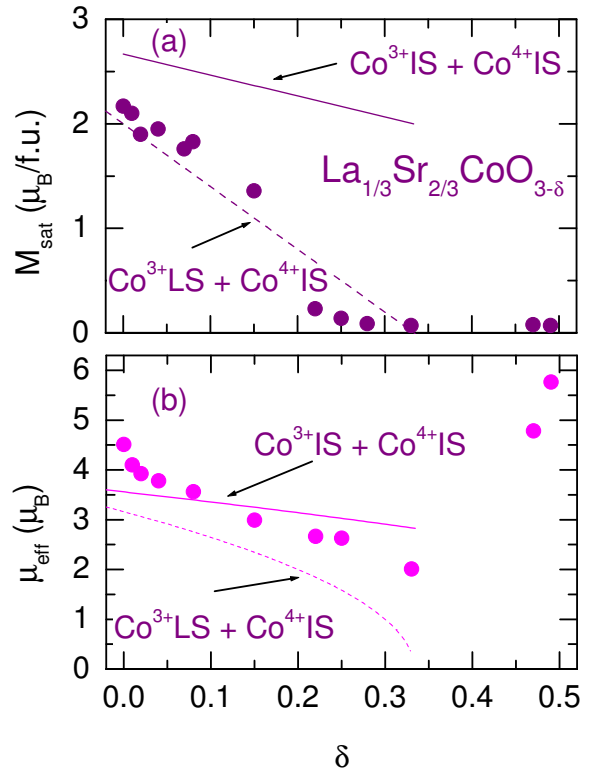}}
 \caption{\label{LSCmom} (color online)
(a) The magnetization in $H=70$~kOe $M_{\rm sat}$ and (b) the effective paramagnetic moments $\mu_{\rm eff}$ for
La$_{1/3}$Sr$_{2/3}$CoO$_{3-\delta}$. Solid and dashed lines illustrate two models of spin states of Co ions (see: text). }
\end{figure}

For $\delta=1/3$ this formal valence of Co is equal to 3+. For smaller $\delta$ we consider a mixture of Co$^{3+}$ and Co$^{4+}$ ions with
the ratio $f_{3+}$Co$^{3+}$:$f_{4+}$Co$^{4+}$, where $f_{3+}=1/3+2\delta$ and $f_{4+}=2/3-2\delta$ are fractions of Co$^{3+}$ and
Co$^{4+}$, respectively. Assuming only ferromagnetic ordering of the cobalt ions, the saturation magnetization for this material can be
described as $M_{\rm sat}=2f_{3+}S_{3+}+2f_{4+}S_{4+}$, where $S_{3+}$ and $S_{4+}$ are spin states for Co$^{3+}$ and Co$^{4+}$,
respectively. Co ions can exhibit several possible spin states: low-spin $t_{2g}^6$ ($S_{3+}=0$), intermediate-spin $t_{2g}^5e_g^1$
($S_{3+}=1$), and high-spin $t_{2g}^4e_g^2$ ($S_{3+}=2$) for Co$^{3+}$  and low-spin $t_{2g}^5$ ($S_{4+}=1/2$), intermediate-spin
$t_{2g}^4e_g^1$ ($S_{4+}=3/2$), and high-spin $t_{2g}^3e_g^2$ ($S_{4+}=5/2$) for  Co$^{4+}$. Our high-field magnetization data, presented
in Fig.~\ref{LSCmom}(a) can be interpreted in the best way by assuming low-spin state for Co$^{3+}$ and intermediate-spin state for
Co$^{4+}$ (Co$^{3+}$LS+Co$^{4+}$IS), which is illustrated in Fig.~\ref{LSCmom}(a) as the dashed line. Only in this case, the low-spin
Co$^{3+}$ ions could account for a fast decrease of the saturation magnetization with $\delta$ in La$_{1/3}$Sr$_{2/3}$CoO$_{3-\delta}$.
The combination of both Co$^{3+}$ and Co$^{4+}$ in the intermediate-spin state [(Co$^{3+}$IS+Co$^{4+}$IS): solid line in
Fig.~\ref{LSCmom}(a)] gives too high values of saturation magnetization.

The effective paramagnetic moment $\mu_{\rm eff}$ is  presented in Fig.~\ref{LSCmom}(b). The spin-only value of the effective paramagnetic
moment for $\delta<1/3$ can be taken as $\mu_{\rm eff}^2=g^2[ f_{3+}S_{3+}(S_{3+}+1)+f_{4+}S_{4+}(S_{4+}+1)]$, where $g=2$ is the
gyromagnetic factor. The $\mu_{\rm eff}$ dependence on $\delta$ is illustrated in Fig.~\ref{LSCmom}(b) as the dashed and solid line for
combinations of Co$^{3+}$LS+Co$^{4+}$IS and Co$^{3+}$IS+Co$^{4+}$IS, respectively. The latter model gives a better approximation to the
$\mu_{\rm eff}$ data. The real values of $\mu_{\rm eff}$ are expected to be higher than the values given by the spin-only model implemented
here, due to the spin-orbit interaction and a nonzero orbital momentum expected for Co$^{4+}$ ions if they are not in the high-spin state.
However, this effect was found to be rather small for other transition metal ions with similar $d$ electron configurations. \cite{Kittel86}
Both $M_{\rm sat}$ and $\mu_{\rm eff}$ can also be reduced by the itinerant character of a fraction of the ferromagnetically coupled
electrons. This effect can be pronounced for $\delta<0.15$ (in the metallic conductivity range) but is expected to be less significant for
larger $\delta>0.15$, (above the metal-insulator transition). The observed different behavior between $M_{\rm sat}$ and $\mu_{\rm eff}$ is
equally striking for both metallic and insulating samples. The magnetic behavior of La$_{1/3}$Sr$_{2/3}$CoO$_{3-\delta}$ can be explained
in a consistent way if we take into account a possible spin transition of Co$^{3+}$ from low-spin to intermediate-spin state similar to
LaCoO$_3$. \cite{Bhide72} In the light of this spin transition, the following spin-state combinations emerge: Co$^{3+}$LS+Co$^{4+}$IS at
$T\leqslant$ 5 K and Co$^{3+}$IS+Co$^{4+}$IS at temperatures above $T_{\rm C}$.

\section{Summary}

In summary, we have synthesized a series of La$_{1/3}$Sr$_{2/3}$CoO$_{3-\delta}$ samples with precisely controlled $\delta=0.00-0.49$. The samples
show significant coupling among the structural, magnetic and transport properties as a function of $\delta$. The stoichiometric material with
$\delta=0.00$ is a cubic ferromagnetic metal with the Curie temperature $T_{\rm C}=274$ K. The increase of $\delta$ to 0.15 is followed by a linear
decrease of $T_{\rm C}$ to $\approx$ 160 K and a metal-insulator transition at the boundary of the cubic structure range. Further increase of
$\delta$ results in formation of tetragonal $2a_p\times 2a_p \times 4a_p$ phase for $\delta\approx 0.25$ and brownmillerite phase for $\delta\approx
0.5$. Those phases are weak ferromagnetic insulators (canted antiferromagnets) with magnetic transitions at $T_{\rm m}=230$ K and 120 K,
respectively.  The $2a_p\times 2a_p \times 4a_p$ phase is $G$-type  antiferromagnetic between 230 K and $\approx$360 K. Magnetic properties of this
system for $\delta<1/3$ can be described in terms of a mixture of Co$^{3+}$ ions in the low-spin state and Co$^{4+}$ ions in the intermediate-spin
state at low temperatures ($T\leqslant$ 5 K) and both Co$^{3+}$ and Co$^{4+}$ ions in the intermediate-spin state at temperatures above $T_{\rm C}$.
For $\delta>1/3$, the $\mu_{\rm eff}$ data suggest a combination of Co$^{2+}$ and Co$^{3+}$ ions, both in the high-spin state with dominating
antiferromagnetic interactions.

\acknowledgments

This work was supported by NSF (DMR-0302617), the U.S. Department of Education, and the U.S. Department of Transportation. At ANL, this
work was also supported by the U.S. Department of Energy, Division of Basic Energy Science-Materials Sciences, under Contract No.
W-31-109-ENG-38 (the operation of IPNS).

 \end{document}